# Spin-orbit-torque and magnetic damping in tailored ferromagnetic bilayers

DongJoon Lee[1,2], JongHyuk Kim[1,3], HeeGyum Park[1,4], Kyung-Jin Lee[2,5], Byeong-Kwon Ju[3], Hyun Cheol Koo[1,2], Byoung-Chul Min[1,4], and OukJae Lee[1]*

[1]*Center for Spintronics, Korea Institute of Science and Technology, Seoul 02792, Korea*

[2]*KU-KIST Graduate School of Converging Science and Technology, Korea University, Seoul 02841, Korea*

[3]*Department of Electrical Engineering, Korea University, Seoul 02841, Korea*

[4]*Division of Nano and Information Technology, KIST school, Korea University of Science and Technology, Seoul, 02792, Korea*

[5]*Department of Materials Science and Engineering, Korea University, Seoul 02841, Korea*

**We study spin-orbit-torque-driven ferromagnetic resonance (FMR) in ferromagnetic (FM) bilayers, consisting of Co and permalloy (Py), sandwiched between Pt and MgO layers. We find that the FM layer in contact with the Pt layer dominantly determines the spin Hall angle, which is consistent with the spin-transparency model. By contrast, the FMR linewidths are considerably influenced not only by the spin-pumping effect across the Pt|FM interface but also by the spin relaxation such as two-magnon scattering at the FM|MgO interface. The Co|MgO interface leads to notably increased FMR linewidths, while the Py|MgO interface has less effect. This different contribution of each interface to the spin Hall angle and dissipation parameter suggests that the stack configuration of Pt|Co|Py|MgO requires less writing energy than Pt|Py|Co|MgO in spin-orbit-torque-driven magnetic switching. Our approach offers a practical tactic to optimize material parameters by engineering either interfaces in contact with the heavy-metal or the oxide layer.**

* Email: ojlee@kist.re.kr

## I. INTRODUCTION

Electrical manipulation of magnetization [1,2,3,4] has been of great importance because of its potential application in low-power and high-speed spintronic devices as well as remarkable scientific findings. More recently, a number of studies have demonstrated that in-plane charge currents generate spin-orbit-torques (SOTs) that can directly switch the magnetization in hetero-structures with a strong spin-orbit coupling (SOC) [2,3,4,5,6]. For the practical use of SOT for magnetic random-access memory (MRAM) applications, it is required to reduce the writing current density ($J_c$) not only for low power consumption and CMOS compatibility [7] but also for enhanced endurance and reliability. The application of a relatively high $J_c$ inevitably generates a substantial amount of Joule heating or electro-migration that could deteriorate magnetic and transport properties in each memory cell, thereby limiting the lifetime of SOT-based MRAMs.

The zero-temperature critical current density ($J_{c0}$) depends on the characteristics of magnetic switching processes as well as many material parameters. For an in-plane magnetization switching (in the rigid-domain approximation) driven by an in-plane anti-damping SOT [1,8],

$$J_{c0} \approx \frac{2e}{\hbar} \frac{\alpha}{\theta_{SH}^{IP}} \left(\frac{4\pi M_{eff}}{2}\right) M_s t_{FM}, \qquad (1)$$

where α is the magnetic damping, $\theta_{SH}^{IP}$ is the conversion efficiency from charge currents to spin currents (having in-plane spin-polarization), $M_s$ is the saturation magnetization, $t_{FM}$ is the magnet thickness, $4\pi M_{eff}$ is the effective out-of-plane demagnetization field ($= 4\pi M_s - 2K_s/M_s t_{FM} > 0$) for an in-plane-magnetized film. For a perpendicular magnetic reversal via coherent rotation by the in-plane anti-damping SOT [9],

$$J_{c0} \approx \frac{2e}{\hbar} \frac{1}{\theta_{SH}^{IP}} \left(\frac{H_k^{eff}}{2}\right) M_s t_{FM}, \tag{2}$$

where $H_k^{eff}$ is the effective out-of-plane field ($= 2K_s/M_s t_{FM} - 4\pi M_s > 0$). In order to achieve a uniform switching of an out-of-plane magnetized film, the lateral dimension of a magnet is limited to less than 30 nm. Otherwise, the magnetic reversal is achieved by the nucleation of reversed domains followed by expansion via domain-wall depinning and propagation [10,11]. Another interesting case is an anti-damping SOT with out-of-plane spin polarization as recently reported [12] in semi-metal WTe2 whose crystal structure has broken lateral mirror symmetry. This type of SOT [12,13] provides a much more efficient pathway for switching of a perpendicular magnet as

$$J_{c0} = \frac{2e}{\hbar} \frac{\alpha}{\theta_{SH}^{OP}} \left(H_k^{eff}\right) M_s t_{FM}, \tag{3}$$

where $\theta_{SH}^{OP}$ is the conversion efficiency from charge currents to spin currents having out-of-plane spin-polarization. A much lower $J_{c0}$ is expected in this case, since $\alpha H_k^{eff} \ll H_k^{eff}/2$ with a typical magnetic damping constant of metallic ferromagnets ($\alpha \approx 0.005 - 0.03$).

In order to minimize $J_c$, hence, it is important to employ a multilayer system exhibiting a large $\theta_{SH}^{eff}$, a small α and a low $H_k^{eff}$. A low resistivity of HM ($\rho_{HM}$) is also desirable to supply more electrons, given the same voltage bias. Unfortunately, there exist trade-offs among those material parameters. Several studies [2,4,14] have demonstrated that a strong SOC material tends to exhibit a larger $\theta_{SH}^{eff}$ as its $\rho_{HM}$ increases. Furthermore, the recent spin-transparency model [15,16] suggests that the simultaneous achievement of a large $\theta_{SH}^{eff}$ and a low α may be improbable. In the model, both the effective magnitude of $\theta_{SH}^{eff}$ and the increased amount of magnetic damping ($\Delta\alpha_{sp}$) owing to the spin-pumping effect [17] are influenced concurrently by the same parameter, namely the interface transparency ($T_{sp}$) of the spin current between a heavy metal (HM) and a FM layer. For instance, Ref. [15] showed $\theta_{SH}^{eff}$(Pt|Co) ≈ 0.11 vs $\theta_{SH}^{eff}$(Pt|Py) ≈ 0.05, but $\Delta\alpha_{sp}$(Pt|Co) ≈ 2·$\Delta\alpha_{sp}$(Pt|Py). This indicates that the increase of $\theta_{SH}^{eff}$ with a higher $T_{sp}$ will be compensated by an increase of α so that a variation of $T_{sp}$ would not be very helpful for lowering $J_c$.

In this paper, we suggest a practical strategy to relieve this issue, using tailored FM bilayers instead of a single FM layer. The bilayers consisting of Co and Py (=$Ni_{79}Fe_{21}$) are sandwiched between Pt and MgO layers and in-plane magnetized. Using spin-torque ferromagnetic resonance (ST-FMR), we investigated how the stacking order and relative thicknesses influence the effective spin Hall angle ($\theta_{SH}^{eff}$), magnetic damping constant (α), inhomogeneous linewidth-broadening ($\Delta H_0$), and effective out-of-plane demagnetization field ($4\pi M_{eff}$). Our results show that (i) the $\theta_{SH}^{eff}$ is mostly determined by the interface material in contact with the Pt layer, and (ii) the α and $\Delta H_0$ are considerably influenced by both interfaces in contact with the HM (Pt) and the oxide layer (MgO).

## II. SAMPLES AND METHODS

We used DC/RF magnetron sputtering to deposit two series of multilayer films with different stacking orders, Pt|Co|Py|MgO or Pt|Py|Co|MgO, on thermally oxidized Si substrates at room temperature (RT). The multilayer structures consist of, from the substrate side, (Co|Py)-series: Ta(1)|Pt(5)|Co($t_{Co}$)|Py(5–$t_{Co}$)|MgO(2)|Ta(2) and (Py|Co)-series: Ta(1)|Pt(5)|Py(5–$t_{Co}$)|Co($t_{Co}$)|MgO(2)|Ta(2) (nominal thickness in nm). The thickness of the Co layer ($t_{Co}$) was varied from 0–5 nm while the total thickness of the FM bilayers was fixed to 5 nm for both series, as illustrated in Fig. 1(a). The magnetic bilayer of Co|Py or Py|Co can be regarded as a single FM unit in our analysis because the total thickness of the bilayer (5 nm) is comparable

to or smaller than the exchange length of each Co or Py layer ($\approx 5 - 10$ nm). The Ta (1 nm) seed layer was employed as a wetting layer, and the MgO(2)|Ta(2) capping layer protects its under-layers and is expected to be fully air-oxidized. The base pressure of the chamber was maintained less than $5 \times 10^{-8}$ Torr and the deposition rates were kept lower than 0.5 Å/s. For the ST-FMR measurement, we patterned the multilayer films into rectangular strips with 15 μm-width ($w$) and 50 μm-length ($l$), using optical lithography and Ar ion milling [Fig. 1(b)]. In a subsequent process step, a waveguide contact made of Ti(10 nm)|Au(100 nm) was defined on top of the samples to apply a microwave electric current to the devices. The samples were not exposed to high temperature (> 120 °C) during the fabrication process as no post-annealing was carried out.

The first step is to characterize the stacking-order dependence of $M_s$ and $H_s^{op}$ (out-of-plane saturation field) for the un-patterned films of the (Co|Py)- and (Py|Co)-series. The $M$–$H$ loop measured by a vibrating sample magnetometer [the inset of Fig. 1(d)] shows that the two FM layers are coupled via a strong exchange energy. Figure 1(c) shows that the $M_s$ of both (Co|Py)- and (Py|Co)- series monotonously increases with increasing $t_{Co}$ because the $M_s$ of Co ($\approx$ 1140 emu/$cm^3$) is larger than that of Py ($\approx$ 640 emu/$cm^3$). The $M_s$ of both (Co|Py)- and (Py|Co)-series are almost identical to each other, thereby indicating that the stacking order of Co|Py or Py|Co on top of the Pt layer has a minor effect on $M_s$. By contrast, as shown in Fig. 1(d), the $H_s^{op}$ has a clear dependence on the stacking order: the $H_s^{op}$ of the (Co|Py)- stack is lower than that of the (Py|Co)-series with the same $t_{Co}$. With the measured $M_s$ and $H_s^{op}$, the net difference of $K_s$ between two stacks is estimated as $\Delta K_s = K_s(Pt|Co|Py|MgO) - K_s(Pt|Py|Co|MgO) \approx 0.6 \pm 0.07$ erg/cm$^2$, which is close to the previously reported [16,18] $K_s \approx 0.8 - 1.1$ erg/cm$^2$ in Pt|Co|MgO. Therefore, the net difference is mostly because of the

Pt|Co interface that has a strong 5d-3d hybridization, resulting in a higher surface anisotropy energy ($K_s$). The Co|Ni and Co|MgO interfaces are known to have non-negligible surface energies, but the contribution to the total anisotropy energy is not that significant in our films.

Next, we systematically investigated their effective spin-Hall angles and magnetic damping constants using ST-FMR [12,14,15,16,19,20] method. The circuit diagram of the ST-FMR measurements is illustrated in Fig. 1(b). A pulsed microwave signal in the range of 4–14 GHz with a nominal output power of 10 dBm was applied to the samples. In the meantime, an external magnetic field-sweep (from -1.8 to +1.8 kOe) was conducted at an angle of 45º within the sample plane. The applied RF current ($I_{rf}$) generates two different types of oscillating SOTs, anti-damping torque ($\tau_{DL}$) $\propto \hat{m} \times (\hat{y} \times \hat{m})$ and field-like torque ($\tau_{FL}$) $\propto \hat{m} \times \hat{y}$, to the magnetization of the adjacent FM bilayer, as well as an Oersted field torque. These torques excite the magnetic precession if frequency and external field satisfy the resonance condition, thereby producing a net oscillation in the anisotropic and spin-Hall magnetoresistances. The mixing of an oscillatory resistance and $I_{rf}$ passing through the FM bilayer generates a finite DC voltage ($V_{mix}$), which is simultaneously detected with a lock-in amplifier connected to the DC port of a bias-tee.

Figures 2(a)–(d) show the representative spectra for the ST-FMR devices with Py(5), Co(5), Py(4)|Co(1), and Co(1)|Py(4) at 8 GHz, exhibiting different resonant positions and linewidths. The single resonance peak observed for all samples indicates that the two FM layers are strongly coupled in-phase. All measured curves are in good agreement with fit to an equation of ST-FMR signal (red curve) consisting of symmetric and asymmetric Lorentzian functions according to:

$$V_{\text{mix}}(H) = S \frac{\Delta H^2}{(H-H_{res})^2+(\Delta H)^2} + A \frac{(H-H_{res})\Delta H}{(H-H_{res})^2+\Delta H^2} \qquad (4)$$

where $S$ ($A$) is the voltage amplitude of the symmetric (asymmetric) Lorentzian function, $H_{\text{res}}$ is the resonance field, and $\Delta H$ is the half linewidth at the half maximum. In Figs. 2(a)-(d), the symmetric (green) and anti-symmetric (blue) parts of the signals are also plotted for comparison. From the ST-FMR theory, the symmetric part ($S$) of the signal is proportional to damping-like SOT acting on the FM magnetization, while the anti-symmetric part ($A$) originates from the sum of field-like SOT and Oersted field torques.

The ST-FMR spectra provide important parameters such as $H_{\text{res}}$, $\Delta H$, $S$ and $A$ that will be used to quantify the magnitudes of $4\pi M_{\text{eff}}$, α, $\Delta H_0$ and $\theta_{SH}^{eff}$. The center frequency of the resonance peak ($f$) follows the Kittel equation, $f = (\gamma/2\pi)\sqrt{H_{\text{res}}(H_{\text{res}} + 4\pi M_{\text{eff}})}$, where $\gamma$ is the gyromagnetic ratio (Fig. 2(e)) and the $4\pi M_{\text{eff}}$ was extracted from a fit to the Kittel equation. The α and $\Delta H_0$ of each sample were derived from $\Delta H = \Delta H_0 + \alpha(2\pi/\gamma)f$ (Fig. 2(f)). The magnitude of $\theta_{SH}^{eff}$ was quantified using the voltage ratio, $S/A$; $\theta_{SH}^{eff} = (S/A)(e 4\pi M_s t_{\text{Pt}} t_{\text{FM}}/\hbar)\sqrt{1 + 4\pi M_{\text{eff}}/H_{\text{res}}}$, where $t_{\text{Pt}}$ is the thickness of Pt (5 nm) and $t_{\text{FM}}$ is the total thickness of the FM bilayers (= 5 nm). This voltage ratio analysis to obtain $\theta_{SH}^{eff}$ should be cautiously applied, since the analysis gives incorrect values if there is a significant magnitude of non-negligible field-like SOT [3,16] or a self-induced Oersted field torque from a non-uniform distribution of current density inside the FM bilayer. We crosschecked that the $\theta_{SH}^{eff}$ obtained in this study coincides with the $\theta_{SH}^{eff}$ extracted by the different analyses, and falls into the category of a proper analysis.

**III. RESULTS: effect of FM stacking order**

The effect of stack order on $4\pi M_{\text{eff}}$, α, $\Delta H_0$ and $\theta_{SH}^{eff}$ has been investigated as a function of Co and Py thicknesses. Several interesting results are illustrated in Fig. 3(a)–(d). Firstly, the magnitude of $4\pi M_{\text{eff}}$ is in good agreement with the value of $H_s^{op}$ in the *M–H* hysteresis; the $4\pi M_{\text{eff}}$ of the (Co|Py)-series is lower than that of the (Py|Co)-series. The difference of interface magnetic anisotropy between Pt|Co|Py|MgO and Pt|Py|Co|MgO stack is estimated from the measured $M_s$ and $4\pi M_{\text{eff}}$ of ST-FMR samples, $\Delta K_s = K_s(Pt|Co|Py|MgO) - K_s(Pt|Py|Co|MgO) \approx 0.55 \pm 0.03$ erg/cm², which is close to the $\Delta K_s \approx 0.6$ erg/cm² obtained with the un-patterned films. The results assure that the ST-FMR samples were not damaged during the device fabrication and reconfirm that the $K_s$ in the Pt|Co|Py|MgO stacks are stronger than that in the Pt|Py|Co|MgO stacks because of the strong $K_s$ at the Pt|Co interface. This difference gives rise to an important consequence in the in-plane magnetization switching driven by the anti-damping SOT, since the *4πM*eff is also one of the key material parameters in determining *J*c0 [see Eq. (1)]. Secondly, for a single FM layer ($t_{\text{Co}} = 0$ or 5 nm), $\theta_{SH}^{eff}(Pt|Co|MgO) \approx 0.14 > \theta_{SH}^{eff}(Pt|Py|MgO) \approx 0.06$ while α(Pt|Co|MgO) ≈ 0.026 > α(Pt|Py|MgO) ≈ 0.018 as shown in Figs. 3(b) and (d). This result is consistent with that expected from the spin-transparency model [15], i.e. a higher spin transparency accompanied with an enhanced magnetic damping.

Thirdly, for FM bilayers (1 nm ≤ $t_{\text{Co}}$ ≤ 4 nm), the measured $\theta_{SH}^{eff}$ are still in agreement with the spin-transparency model in which the interface between the FM and Pt layer mostly determines $\theta_{SH}^{eff}$. By contrast, the measured α of Pt|FM-bilayers shows some inconsistency with the spin transparency model. Figure 3(b) shows that the α of Pt(5)|Py(5-$t_{\text{Co}}$)|Co($t_{\text{Co}}$)|MgO(2) series is more or less the same as the α of Pt(5)|Co(5)|MgO(2), ≈ 0.026,

and the α of Pt(5)|Co($t_{Co}$)|Py(5- $t_{Co}$)|MgO(2) is very close to the Pt(5)|Py(5)|MgO(2), ≈ 0.018. Both results show a weak $t_{Co}$-dependence, indicating that the magnitude of α is more significantly influenced by the material adjacent to MgO or FM|MgO interface than the material adjacent to Pt or the Pt|FM interface. This is in discord with the spin-transparency model where the enhanced damping is mainly determined by the spin-pumping effect occurring through the Pt|FM interface. Instead, we found that the magnitude of $α$ has a relation to the magnitude of $ΔH_0$, which is another magnetic relaxation parameter associated with the magnetic inhomogeneity of a sample. Fig. 3(c) shows that $ΔH_0$ of the bilayers also significantly depends on the FM material next to the MgO layer rather than on the FM material interfacing the Pt layer. The magnitude of $ΔH_o$ was four to five times higher in the (Py|Co)-devices than that of the (Co|Py)-devices since we observed $ΔH_o$ (Pt|Py|Co|MgO) ≈ 62 ± 4 $Oe$ > $ΔH_o$ (Pt|Co|Py|MgO) ≈ 13 ± 6 $Oe$ for 1 ≤ $t_{Co}$ ≤ 4 $nm$.

The results shown above clearly suggest that the FM|MgO interface considerably contributes to the magnitudes of α and $ΔH_0$. A remaining question is how large portion of magnetic damping in the FM bilayers originates from the FM|MgO and the Pt|FM interface respectively. In order to extract the contribution of the Pt|FM interface, we have conducted an additional ST-FMR experiment for almost identical multilayer stacks but without a Ta(1)|Pt(5) buffer layer. The measured α and $ΔH_0$ are also plotted in Figs. 3 (b) and (c). The ST-FMR signal in the devices was barely measurable, and the resonant peaks were distinguishable only when an inhomogeneous current density existed in the magnetic bilayer that can be excited by itself. Furthermore, some of the devices did not exhibit any discernible spectra within our measurement resolution. Hence, the result that we were able to obtain is the averaged magnitude of α, α(Co|Py|MgO) ≈ 0.01 and α(Py|Co|MgO) ≈ 0.012. These are somewhat

lower than α(Pt|Co|Py|MgO) ≈ 0.018 and α(Pt|Py|Co|MgO) ≈ 0.03 respectively, thereby confirming the increase of the damping due to the Pt|FM interface as well.

If this enhanced damping solely arises from the spin-pumping effect, the Pt|(Co|Py) interface has the effective spin-mixing conductance: $g_{\text{eff}}^{\uparrow\downarrow} = \frac{4\pi M_s t_{\text{FM}}}{\gamma \hbar}[\alpha(Pt|Co|Py|MgO) - \alpha(Co|Py|MgO)] \approx 21 \pm 7 \; nm^{-2}$. Thus, the spin-transparency is $T_{sp} \approx \frac{2(e^2/h)g_{\text{eff}}^{\uparrow\downarrow}}{1/\lambda_{\text{Pt}}\rho_{\text{Pt}}} \approx 0.72 \pm 0.24$, where $\lambda_{\text{Pt}}$ is the spin-diffusion length in Pt (if using $\lambda_{\text{Pt}} \approx 1 \; nm$ [15,16]) and $\rho_{Pt}$ is the electrical resistivity of Pt ($\approx 45 \; \mu\Omega \cdot cm$) on top of the Ta(1) buffer layer. The calculated $T_{sp}$(Pt|(Co|Py)) gives the intrinsic magnitude of the spin-Hall angle ($\theta_{SH}^{int}$) in Pt as large as 0.19 ($\because \theta_{SH}^{eff} = T_{sp} \cdot \theta_{SH}^{int}$), where our $T_{sp}$ and $\theta_{SH}^{int}$ are in good agreement with the previously reported [15] $T_{sp}$ ($\approx 0.65$) at the interface of Pt|Co and $\theta_{SH}^{int}$ of Pt ($\approx 0.17$) although we had quantitatively different values in $g_{\text{eff}}^{\uparrow\downarrow}$ and $\rho_{\text{Pt}}$. The same analysis is applied to obtain of $g_{\text{eff}}^{\uparrow\downarrow}$ and $T_{\text{sp}}$ of the Pt|(Py|Co) interface, but this gives rise to an unphysical consequence: $g_{\text{eff}}^{\uparrow\downarrow} \approx 32 \pm 10 \; nm^{-2}$ and consequently $T_{\text{sp}} \approx 1.1 \pm 0.36$. The obtained $g_{\text{eff}}^{\uparrow\downarrow}$ remains consistently in the range of previous reports: $g_{\text{eff}}^{\uparrow\downarrow}(Pt|Py) \approx 20 - 40 \; nm^{-2}$, but the $T_{sp}$ larger than unity is an unphysical fallout.

We note that the estimation of $T_{sp}$ is uncertain at the moment, mainly due to the variation in the reported values of $\lambda_{Pt}$ by more than one order of magnitude, ranging from 1 to 11 nm, so that the used value ($\lambda_{\text{Pt}} \approx 1$ nm) above is possibly to be incorrect. A recent work [21] has reported an important relation, $\lambda_{Pt} \propto \rho_{Pt}^{-1}$, in order to reconcile such discrepancies, with the assumption that Elliott-Yafet (EY) scattering mechanism dominates the spin relaxation in Pt at RT. The scenario is quite reasonable because another work [22] has confirmed that EY scattering is dominant in Pt at RT while D'yakonov-Perel' (DP) relaxation dominates at cryogenic temperatures. The reported spin-resistance ($r_{s,Pt} = \lambda_{Pt} \cdot \rho_{Pt}$) of Pt is still scattered over a wide range of magnitude: $r_{s,Pt} \approx 0.6$ f$\Omega \cdot$m$^2$ [23], $r_{s,Pt} \approx 0.77$ f$\Omega \cdot$m$^2$ [21], and $r_{s,Pt} \approx 2.6$ f$\Omega \cdot$m$^2$ [24]. Given the relation $\lambda_{Pt} \propto \rho_{Pt}^{-1}$, the expected $\lambda_{Pt}$ from our Pt films could be, respectively, 1.3 nm, 1.7 nm, and 5.8 nm, resulting in the $T_{SP}$ (Pt|(Co|Py)) ≈ 0.94, 1.2, and 4.2. Even for the case of $T_{SP} \approx 0.94$, the corrected $T_{SP}$ would be larger than one if taking account of spin-memory-loss (SML) at the interface [23]. The estimated values of $T_{SP}$ are unphysical regardless of its variation, since it is greater than unity for all of the cases.

We could understand such unphysical results in two different ways. The first case happens if the spin current is generated at the Pt|FM interface rather than in bulk Pt [25,26]. The origin of spin current is still under significant debate in the spintronic community so its conclusion is beyond our scope. The other is the case when the magnitude of $g_{\text{eff}}^{\uparrow\downarrow}$ is overestimated for Pt|FM bilayers. By now, most of experimental works for $g_{\text{eff}}^{\uparrow\downarrow}$ have assumed that the enhanced damping with the Pt-interface is *solely* due to the spin-pumping effect, and thereby calculated the value of $g_{\text{eff}}^{\uparrow\downarrow}$ from the difference in the magnetic damping between two stacks with and without Pt-interface. However, there might exist additional dissipation channels for the spin-dynamics such as interfacial spin-flip scattering [27,28] or two-magnon scattering (TMS) [29,30], originating from the Pt|FM interface. Ref. [23] has studied the SML effect at the Pt|Co interface in which the transmission of spin-current is reduced at the interface due to the interfacial spin-flip scattering. The origin of SML effect has been attributed to the magnetic proximity effect in Pt [23] or to the development of magnons due to the interfacial (non-collinear) Dzyaloshinskii-Moriya interaction [31]. In the same way, such effects should influence on the dissipation of spin dynamics as well, resulting in the enhancement of magnetic damping.

The possibility of additional relaxation channels at the Co|MgO interface are also revealed by the increased inhomogeneity in Fig. 3(c), where $\Delta H_0$ (Pt|Py|Co|MgO) > $\Delta H_0$ (Pt|Co|Py|MgO). The Co in contact with MgO leads to broadening of resonance peak and, in other words, additional spin relaxations. Moreover, the (Co|Py)-devices without a Ta(1)|Pt(5) buffer, where the Co layer directly interfaces to the $SiO_x$ layer, exhibits a larger $\Delta H_0$ than that in the (Co|Py)-devices with the Ta/Pt buffer layer (see Fig. 3(c)). This implies that the presence of Co adjacent to oxides or a possible formation of interfacial Co-oxide gives rise to increased magnetic inhomogeneity and consequently enhanced magnetic damping as well.

An important consequence of our results described so far is that, even with the same material combinations of Co and Py, the stacking order influences the $4\pi M_{\text{eff}}$, α, $\Delta H_0$ and $\theta_{\text{SH}}$, which in turn affect the writing energy required for switching the Co and Py magnetization. Combining the measured parameters in Figs. 3(a)–(d), we used Eq. (1) to calculate the magnitude of $J_{c0}$ as a function of $t_{Co}$ for the (Co|Py)- and (Py|Co)-series (see Fig. 3(e)). The (Co|Py)-series have consistently smaller $J_{c0}$'s than the (Py|Co)-series for the same $t_{Co}$ because of combined effects: the former group has smaller $\alpha$ (and $\Delta H_0$) and $4\pi M_{\text{eff}}$ but larger $\theta_{SH}^{eff}$ than the latter. The result suggests a possibility that can lead to a reduction of $J_{c0}$ by optimizing the magnetic stack order and the interfaces in contact with the HM and the oxide layer.

We note that our bilayer approach can be applicable to a perpendicularly magnetized system as well. Most of experimental studies have utilized magnetic layers of 1-2 nm thickness for SOT-driven magnetic switching, because in the thickness ranges the interfacial energy becomes sufficient to compensate the volume demagnetization energy, so as to make the system perpendicularly magnetized or to lower the $4\pi M_{\text{eff}}$ for an in-plane magnetized system. Furthermore, the magnitude of SOT becomes enough to reverse the polarity of magnet since it is proportional to $(M_S \times t_{FM})^{-1}$. Naturally one can control $4\pi M_{\text{eff}}$ or $H_k^{eff}$ of the Co|Py bilayers by optimizing their relative thicknesses within total 1-2 nm, while minimizing the $\alpha$ and maximizing the $\theta_{SH}^{eff}$ from the stacking order and interface engineering. We also expect that the interfacial contributions will be more significant as the total thickness decreases and the FM bilayer becomes perpendicularly magnetized.

**IV. DISCUSSION: contribution of FM|MgO interface on magnetic damping**

As mentioned previously, we presume that there exists a strong correlation between α and $\Delta H_0$. To further clarify our speculation, we have conducted additional ST-FMR experiments for the stacks comprising a FM dusting layer at the interface adjacent to the MgO layer. The devices in this study have FM multilayers consisting of: (Co2|Py2|Co)-series: Ta(1)|Pt(5)|Co(2)|Py(2)|Co($t_{Co}$)|MgO(2)|Ta(2); (Py2|Co2|Py)-series: Ta(1)|Pt(5)|Py(2)|Co(2)|Py($t_{Py}$)|MgO(2)|Ta(2); (Co4|Py)-series: Ta(1)|Pt(5)|Co(4)|Py($t_{Py}$)|MgO(2)/Ta(2); and (Py4|Co)-series: Ta(1)|Pt(5)|Py(4)|Co($t_{Co}$)|MgO(2)|Ta(2), between Pt and MgO layers. Both $t_{Co}$ and $t_{Py}$ were varied from 0–1 nm and the thickness of the FM single or bilayer, which is adjacent to Pt, was fixed to 4 nm, as illustrated in Figs. 4(a) and 4(b).

Figures 4 (c)–(f) show the obtained $\alpha$ and $\Delta H_0$, as functions of $t_{Co}$ or $t_{Py}$, in which one can see apparent effects of the FM dusting layer. First, the insertion of Py dusting layer in between Co and MgO leads to a reduction of $\alpha$ and $\Delta H_0$. For example, as we increase the thickness of Py dusting layer ($t_{Py}$) from 0 nm to 1 nm in the devices of the (Py2|Co2|Py)- and (Co4|Py)-series (solid shape in the figures), the $\alpha$ is gradually reduced from 0.034 to 0.016 for the Py2|Co2|Py($t_{Py}$) series, and from 0.029 to 0.015 for the Co4|Py($t_{Py}$) series (Fig. 4(c) and (d)); the $\Delta H_0$ is reduced from 90 Oe to 10 Oe for the Py2|Co2|Py($t_{Py}$) series and from 70 Oe to 20 Oe for the Co4|Py($t_{Py}$) series (Fig. 4(e) and (f)). Second, the insertion of Co dusting layer in between Py and MgO leads to an enhancement of $\alpha$ and $\Delta H_0$. For instance, as we increase the thickness of Co dusting layer ($t_{Co}$) from 0 nm to 1 nm in the (Co2|Py2|Co)- and (Py4|Co)-series (cross shape in Figs. 4(a) and (b)), the $\alpha$ is gradually enhanced from 0.02 to 0.029 for the Co2|Py2|Co($t_{Co}$) series, and from 0.022 to 0.029 for the Py4|Co($t_{Co}$) series (Fig. 4(c) and (d)); the $\Delta H_0$ is increased from 20 Oe to 30 Oe in the Co2|Py2|Co($t_{Co}$) series, and from 2 Oe to 60 Oe in the Py4|Co($t_{Co}$) series (Fig. 4(e) and (f)).

Our experimental results described above clearly demonstrate that α and $\Delta H_0$ are strongly correlated. Figure. 5(a) shows the plot of $\alpha$ $vs$ $\Delta H_0$ determined from all our ST-FMR devices having FM bi- or tri-layers. For the samples with a Py|MgO interface (solid symbols), $\Delta H_0$ is mostly distributed below 20 Oe, and the $\alpha$ remains as small as 0.02. For the samples with a Co|MgO interface (open symbols with a cross in the figure), the $\Delta H_0$ is distributed in a wide range from 0 to 60 Oe, and the $\alpha$ is enhanced quasi-linearly with increasing $\Delta H_0$. The increase of $\Delta H_0$ has been known to be related to the samples' (magnetic) inhomogeneity. The enhanced $\alpha$ associates with the increased $\Delta H_0$ definitely indicating that a new magnetic relaxation channel is developed at the Co|MgO interface.

To the best of our knowledge, there is no theoretical model that directly connects $\Delta H_0$ to α. An important clue observed from our experiment is fact that both $\Delta H_0$ and α are increased with decreasing thickness ($t_{FM}$) of an ultrathin FM film; i.e., α and $\Delta H_0 \propto 1/t_{FM}^n$ in general. In order to understand this, we have checked several mechanisms such as spin-pumping, spin scattering, and TMS that might be account for the increased $\Delta H_0$ and α. First, the spin-pumping effect [17] occurs when a FM is interfacing to a strong spin-scatterer (e.g., Pt), but this is not the case for our Co oxide interface. Second, the spin scattering can be increased with decreasing $t_{FM}$ because the surface contribution [27,28] significantly increases the electron scattering rate ($1/\tau_s$) and consequently the spin-flip scattering rate ($1/\tau_{sf}$). However, this mechanism is not in accordance with our observation in which the measured resistivities of FM layers from all ST-FMR devices exhibit no clear correlation to α (see Fig. 5(b)). Third, a probable mechanism that might be related to our experiment is the TMS [29,30], which contributes to the magnetic damping of ultrathin ferromagnetic films in which the uniform mode ($k = 0$) is excited by ST-FMR scatters into degenerated magnons ($k \neq 0$) due to surface roughness or defects, which is strongly related to the magnetic disorder.

The *antiferromagnetic* (AF) formation of interfacial magnetic oxides (CoO, NiO and $Fe_2O_3$) can be attributed to the source of dynamic non-uniformity and enhanced magnetic damping via TMS process. The interfacial AF layer can change the magnetic behavior of the FM with the introduction of extra anisotropies via exchange-bias effects. A distribution of grains in the AF-oxide induces anisotropy fluctuation in random directions. Thereby the AF-oxide can open an additional relaxation pathway and spatially non-uniform dynamics during the magnetic precession through the exchange coupling to the fluctuating spins of AF-oxide and by the slow dragging of AF-oxide domains.

We presume, however, that it is unlikely that possible interfacial oxides in our stacks have an antiferromagnetic order at RT. The exchange bias becomes effective when the AF-oxide becomes block magnetically, i.e. when the temperature is below the blocking temperature ($T_B$). The $T_B$'s of bulk CoO and NiO are known to be ≈ 290 K and ≈ 470 K respectively [32], but the ones in ultrathin film are often very different from the bulk values for AF-oxides [33]. For instance, 2.5 nm-thick NiO has $T_B$ ≈ 200 K [34] and in general the $T_B$ decreases with decreasing the thickness of the AF-oxide. Our interfacial CoO, NiO, and $Fe_2O_3$ layers should have much lower $T_B$'s than RT because their effective thicknesses are expectedly ~0.6-0.8 nm, estimated from the observation in Fig. 4 (c)-(d) at which the damping becomes saturated. Moreover, the $T_B$ of CoO is expected to be even lower than the one of NiO, so that the Py-MgO interface should have more substantial interfacial AF-oxide than the Co-MgO interface. This is in opposition to our observation.

Nevertheless, it would be useful study FMR-linewidths as functions of measurement and annealing temperatures. The strength of exchange bias from the interfacial AF-oxide is strongly dependent on the microstructural properties of both FM and MgO such as grain size distribution, structural defect, interfacial roughness, enthalpy of formation and variation in chemical composition. Advanced interface chemical analysis such as X-ray photoemission spectroscopy (XPS), electron energy loss spectroscopy or polarized-neutron reflectometry can be useful tools to investigate the microstructural properties of both FM and MgO as well. Furthermore, in order to confirm the development of TMS at the Co|MgO interface, further experiments are required such as the measurement of $\Delta H$ as functions of wide ranges of angle and frequency [35].The results will provide better explanation about the physical origins of linewidth broadening at the Co|MgO interface.

The bilayer approach might give one possible strategy that can reduce the switching current in the STT/SOT-MRAM technology. In a typical MRAM configuration, CoFeB-alloy has been utilized as a magnetic free layer that interfaces to MgO-tunnel barrier. We note that Py with MgO tunnel barriers does not provide a good tunnel magneto-resistance (TMR), which makes the proposed configuration be ineffective in the electrical readout of the magnetic change. If the oxidation of Co plays an important role in the interfacial enhancement of damping, the damping could be reduced by the insertion of a magnetic dusting layer, for instance Fe or FeB, between CoFeB and MgO layers. This may be an alternative way for lowering the write energy as long as the interface engineering does not damage other parameters, such as TMR, thermal stability and spin-polarization. We believe our results suggest a practical strategy in the optimization of SOT and magnetic damping by engineering both interfaces in contact with the HM and the tunnel-barrier layer, for achieving more energy-efficient MRAM.

## V. SUMMARY

In summary, we utilized Co/Py FM-bilayers, which are in-plane magnetized and sandwiched between Pt and MgO layers, as test-beds for understanding the roles of interfaces in the spin transparency and understanding the interfacial contributions on the magnetic damping. Our results demonstrated that the magnitude of $\theta_{SH}^{eff}$ is mostly determined by the interfacial material in contact with the Pt layer, which is consistent with the spin transparency model. By contrast, the magnetic relaxation parameters, $\alpha$ and $\Delta H_0$, are substantially influenced not only by the Pt|FM interface but also by the spin relaxation such as TMS at the FM|MgO interface. Both α and $\Delta H_0$ are significantly increased at the Co|MgO interface probably via increased TMS processes whereas such increases are negligible at the Py|MgO interface. In order to achieve a low writing current density in SOT-driven magnetic switching, the multilayer configuration of Pt|Co|Py|MgO has more preferable material parameters than the stack of Pt|Py|Co|MgO.

**Acknowledgements**

This work was supported by the National Research Council of Science & Technology (NST) grant (No. CAP-16-01-KIST) and the KIST Institutional Program (2E28190). K. -J. L. was supported by the National Research Foundation of Korea (NRF) [NRF-2017R1A2B2006119] and KU-KIST School Project. K. -J. L. acknowledges the KIST Institutional Program (Project No. 2V05750)

**Figure Captions**

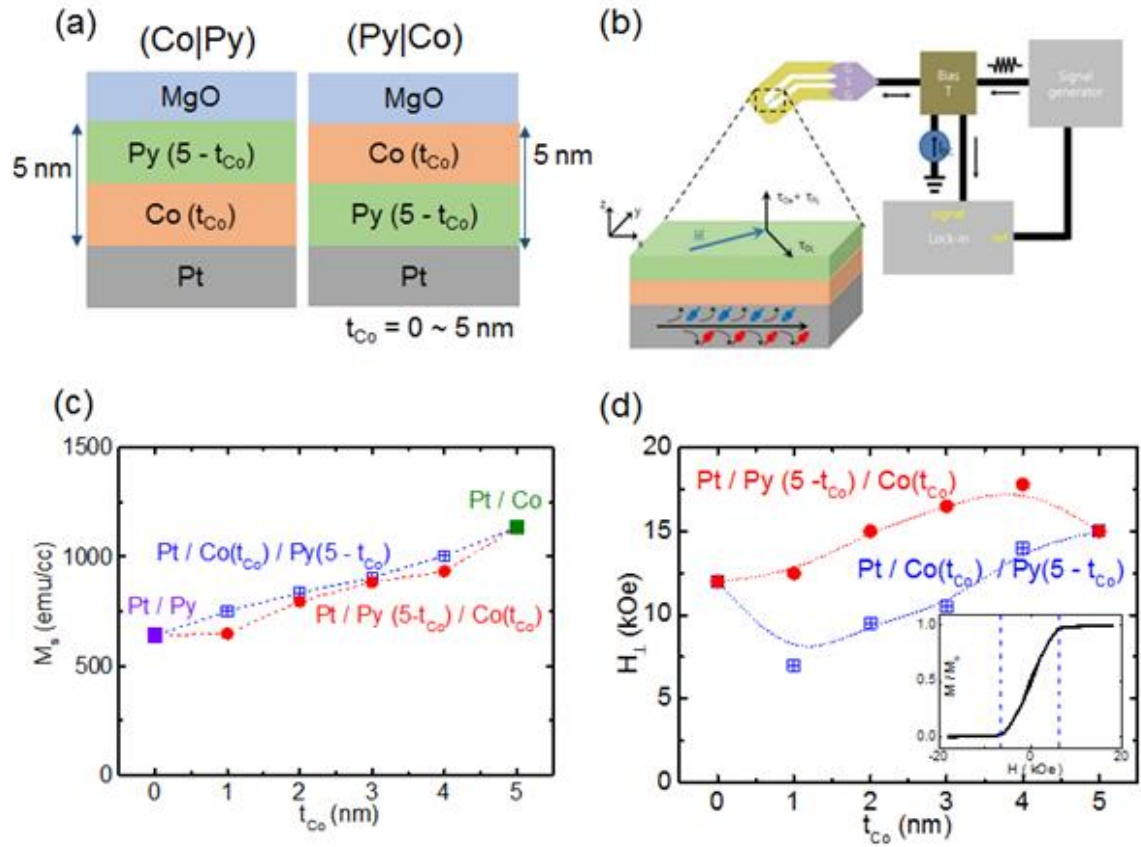

**FIG. 1.** (a) Multilayer structures of (Co|Py)- and (Py|Co)-series. The Co layer thickness ($t_{Co}$) was varied from 0 to 5 nm while the total thickness of the FM bilayers was fixed to 5 nm. (b) Schematic of ST-FMR measurement. (c) Saturated magnetization ($M_S$) and (d) $H_s^{op}$ (out-of-plane saturation field) for un-patterned films of (Co|Py)- and (Py|Co)-series.

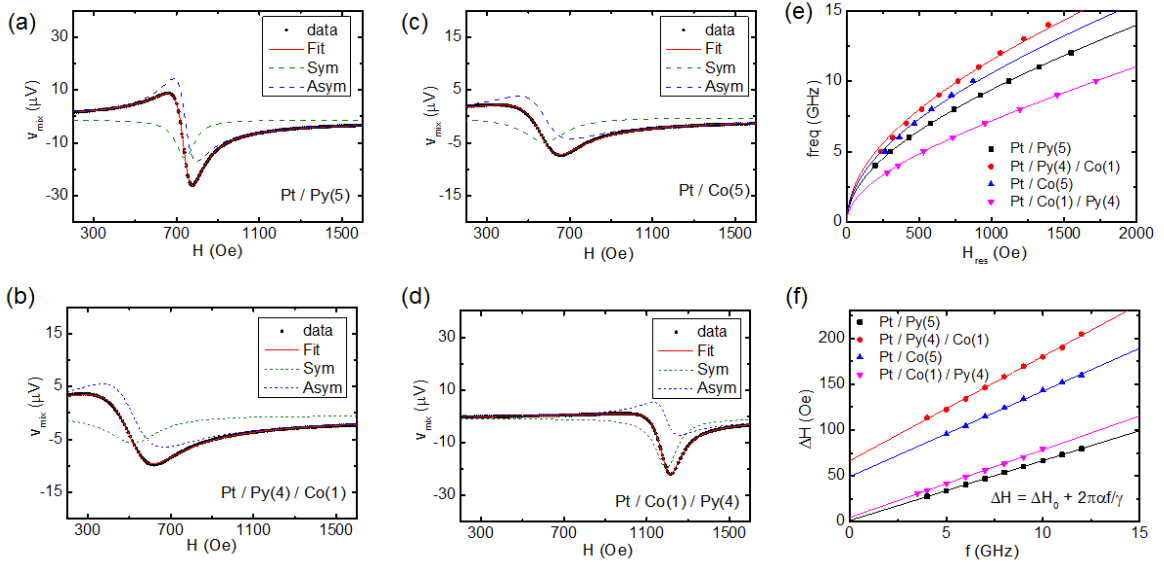

**FIG. 2.** (a)-(d) Representative spectra for ST-FMR devices with Py(5), Co(5), Py(4)|Co(1), and Co(1)|Py(4) at 8 GHz, exhibiting different resonant positions and linewidths. Red curves are fit to Eq. (1) (red curve). The symmetric (green) and anti-symmetric (blue) parts of the signals are also plotted. (e) Obtained $H_\text{res}$ and (f) $\Delta H$ as a functions of *f* along with fitting curves.

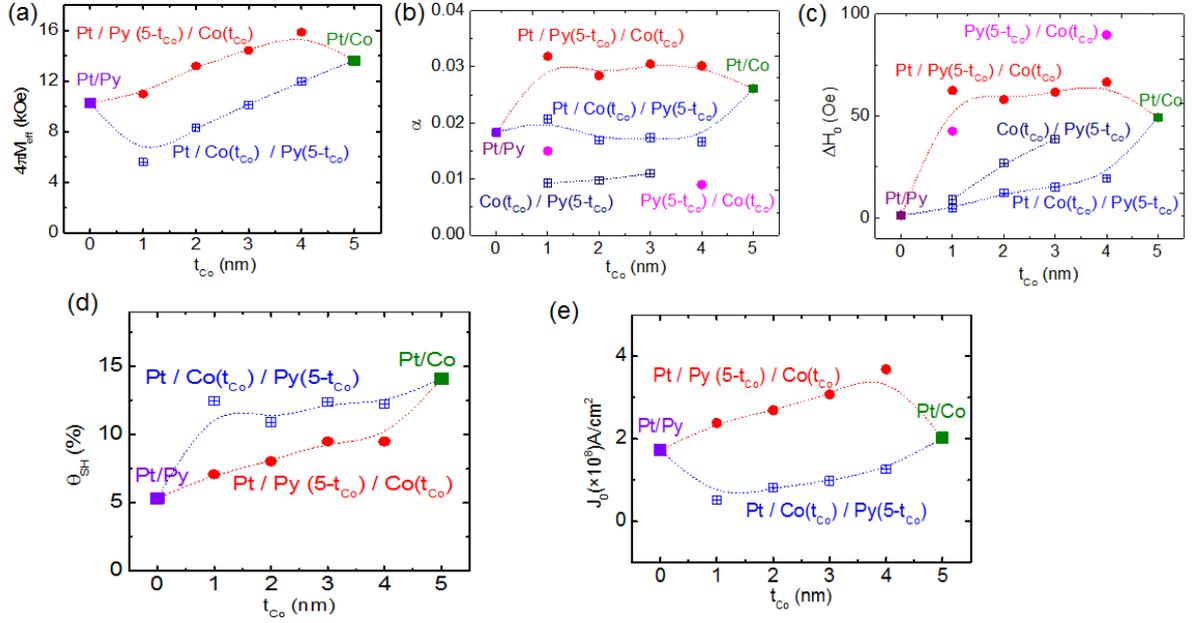

**FIG. 3.** Measurement results of (a) $4\pi M_{\text{eff}}$, (b) $\alpha$, (c) $\Delta H_0$ and (d) $\theta_{SH}^{eff}$ as functions of stacking order and thickness of Co and Py. (e) Calculated $J_{c0}$ as a function of $t_{\text{Co}}$ for (Co|Py)- and (Py|Co)-series by using Eq. (1) with the measured parameters in (a)–(d).

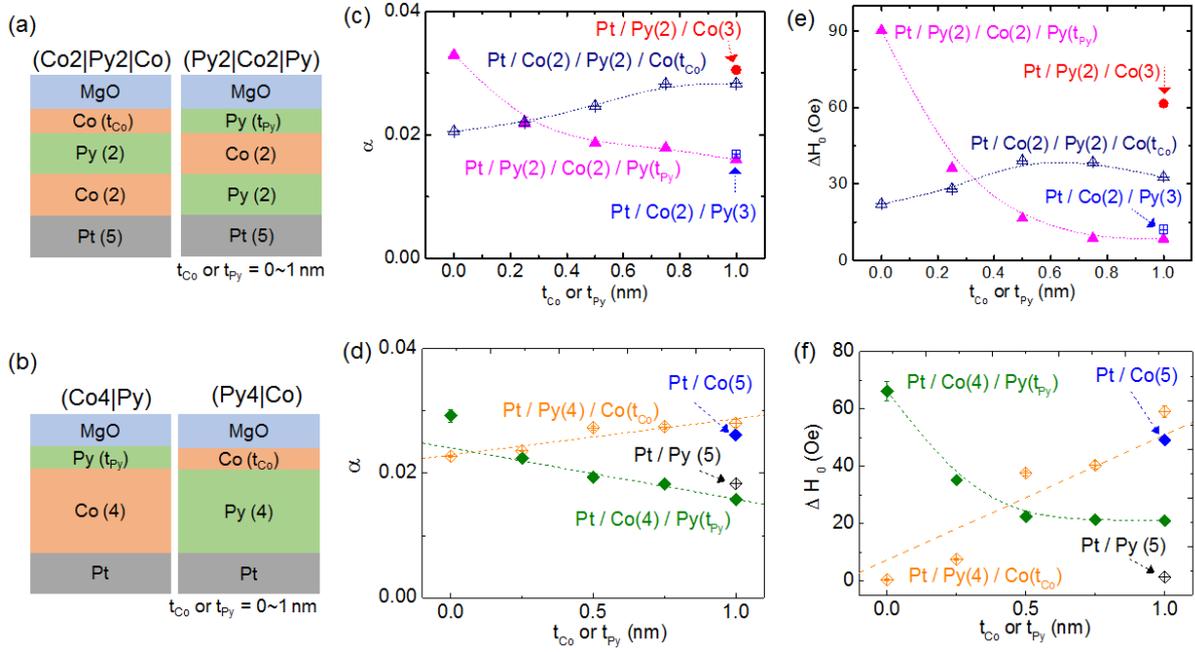

**FIG. 4.** Multilayer structure of (a) (Co2|Py2|Co)-, (Py2|Co2|Py)-, (b) (Co4|Py)-, and (Py4|Co)-series. Additional FM dusting layer was inserted at the interface of FM|MgO and its thickness was varied from 0 to 1 nm. The thickness of the initial FM layer was fixed to 4 nm. (c)–(f) Obtained $\alpha$ and $\Delta H_0$ as functions of $t_{Co}$ or $t_{Py}$.

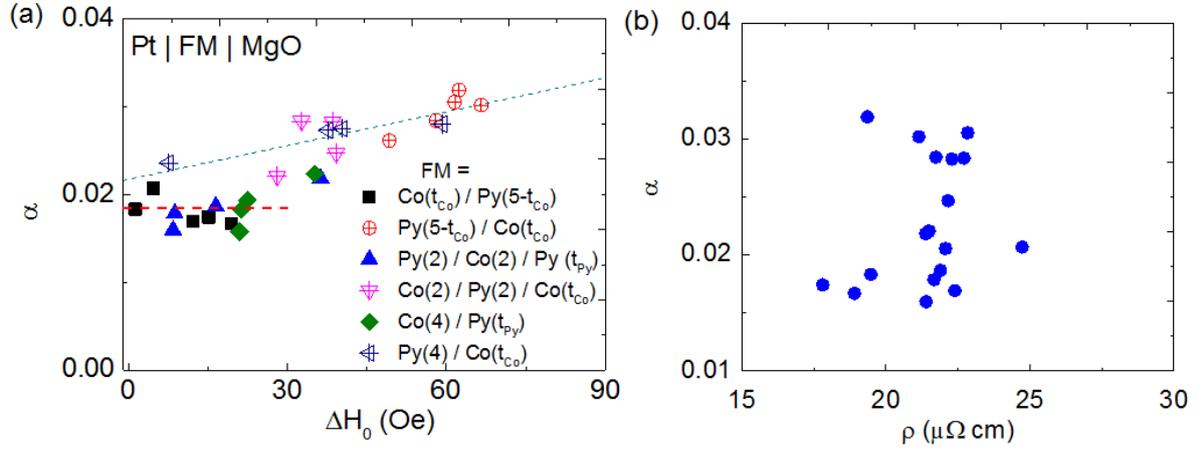

**FIG. 5.** (a) Plot of $\alpha$ vs $\Delta H_0$ of devices having FM bi- or tri-layers. Data for the samples with the Co|MgO interface are represented by cross shapes, and data for devices with the Py|MgO interface is by solid shapes. (b) Plot of $\rho$ (the resistivity of FM bi- or tri-layers) vs $\alpha$.